\documentclass{aa}

\usepackage{graphicx,times}
\usepackage{graphics}
\def\simlt {\lower.5ex\hbox{$\; \buildrel < \over \sim \;$}} 
\def\simgt{\lower.5ex\hbox{$\; \buildrel > \over \sim \;$}} 
\def\msun {\hbox{M$_{\odot}$}}

\def\arcmin {$^\prime$} 
\def\arcsec{$^{\prime\prime}$}

\begin{document}

   \thesaurus{06     
              (03.11.1;  
               16.06.1;  
               19.06.1;  
               19.37.1;  
               19.53.1;  
               19.63.1)} 
   \title{X-ray Spectroscopy of the Cluster of Galaxies Abell 1795 with XMM-Newton}

   \subtitle{}

   \author{T.Tamura \inst{1}
          \and
	J.S. Kaastra \inst{1}
	\and
	J. R. Peterson \inst{2}	
	\and
	F. Paerels \inst{2}
	\and
	J. P.D. Mittaz \inst{3}	
	\and	
	S. P. Trudolyubov \inst{4}	
	\and	
	G. Stewart \inst{5}
	\and
	A.C. Fabian \inst{6}
	\and
	R.F. Mushotzky \inst{7}
	\and
	D. H. Lumb \inst{8}
	\and
	Y. Ikebe \inst{9}
	}
	   \offprints{T.Tamura}
	\mail{T.Tamura@sron.nl}

\institute{ SRON Laboratory for Space Research
              Sorbonnelaan 2, 3584 CA Utrecht, The Nether\-lands 
              \and
              Astrophysics Laboratory, Columbia University, 
              550 West 120th Street, New York, NY 10027, USA
              \and
		Department of Space and Climate Physics,
		University College London, Mullard Space Science Laboratory,
		Holmbury St. Mary, Surrey, U.K.
		\and
		NIS-2, Los Alamos National Laboratory, Los Alamos, NM 87545, USA
		\and
		Department of Physics and Astronomy, The University of Leicester, Leicester LE1, UK
		\and
		Institute of Astronomy, University of Cambridge, Madingley Road, Cambridge, CB3 0HA, UK
		\and
		NASA Goddard Space Flight Center , Code 662, Greenbelt, Maryland 20771, USA
		\and
		 Astrophysics Div., European Space Agency, ESTEC, Postbus299,
		2200AG Noordwijk, The Nether\-lands 
		\and
	              Max Planck Institut f\"ur Extraterrestrische Physik,
	Postfach 1312, 85741 Garching, Germany
}

   \date{Received 3 Oct. 2000; accepted 18 Oct. 2000}

   \maketitle

\begin{abstract}
The initial results from XMM-Newton observations of the rich cluster of galaxies Abell 1795 are presented.
The spatially-resolved X-ray spectra taken by the European Photon Imaging Cameras (EPIC) show a temperature drop at a radius of $\sim 200$ kpc from the cluster center, 
indicating that the ICM is cooling.
Both the EPIC and the Reflection Grating Spectrometers (RGS) spectra extracted from the cluster center 
can be described by an isothermal model with a temperature of $\sim 4$ keV.
The volume emission measure of any cool component ($<1$~keV)
is less than a few \% of the hot component at the cluster center.
A strong \ion{O}{viii} Lyman~$\alpha$ line was detected with the RGS from the cluster core.
The O abundance and its ratio to Fe at the cluster center is 0.2--0.5 and 0.5--1.5 times the solar value, 
respectively.

   \end{abstract}

%

\section{Introduction}
X-ray observations have revealed 
that the radiative cooling time of the intra-cluster-medium (ICM) around the central galaxy of many clusters is much less than the cluster age (e.g. Edge, Stewart, and Fabian \cite{edge}).
This indicates that cooling of the ICM could strongly affect the cluster thermal structure 
(e.g. Fabian \cite{fabian-cf}). 
Although pre-Chandra/XMM observations established the presence of plasma 
near the cluster center considerably cooler than that found further out in the cluster (e.g. Allen et al. ~\cite{allenetal}; Makishima et al. \cite{makishima}),
the detailed spatial and thermal structure of this cool component, and hence its nature, 
has not yet been fully understood. 
In addition the measurement of the elemental abundances in this cooler gas provides critical information concerning its formation and evolution.


Here we report the first results from the XMM (Jansen et al. \cite{jansen}) observations of a rich cluster of galaxies, 
Abell 1795 (A1795 for short).
This cluster, at a redshift of 0.0631, 
is one of the best targets for spectral study of the cluster center with XMM.
This is because of its peaked X-ray emission and a cool (2--4 keV) component at the cluster center (Briel and Henry \cite{briel}; Fabian et al. \cite{fabian}; Fabian et al. \cite{fabian-chandra}).
Based on the the deprojection of the X-ray image a strong cooling flow with a mass deposition rate, $\dot{M}$, of $\sim 500$~\msun$/$yr
was reported (Edge et al. \cite{edge}; Allen et al. \cite{allenetal}).
The ASCA spectra showed a smaller level of the cooling flow with 
$\dot{M}$ of $\sim 250$~\msun$/$yr (Allen et al. \cite{allenetal}).

We have measured the temperature structure and metal abundances of the central region
based on X-ray spectra with high spectral resolution.
Throughout this $Letter$,
we assume the Hubble constant to be $H_0 = 50$ km s$^{-1}$Mpc$^{-1}$
and use the 90\% confidence level unless stated otherwise. 
One arc-minute corresponds to 110 kpc.

\section{Observations}
XMM observations of A1795 were performed on 2000 June 26 during the performance verification phase.
The three EPICs (Turner et al. \cite{turner})  
were operated in the full window mode with the thin filter.
The RGS (den Herder et al. \cite{herder}) were operated in the spectroscopy mode.
The RGS dispersion axis is oriented along an axis of about $-70$ degrees (North to East).
We report on simultaneous UV/optical observations with the Optical Monitor in another $Letter$ (Mittaz et al. \cite{mittaz}).

\section{Analysis}
Using the EPIC spatially-resolved spectra we have measured large-scale properties of the cluster.
Then, utilizing the high resolution RGS spectra we constrain the temperature structure in the cluster core.

For basic data processing, we used the development version of the Science Analysis System (SAS) for all data.
The EPIC and RGS events were screened by rejecting the high background periods.
Useful exposure times for EPIC/PN, EPIC/MOS and RGS are 25 ksec, 34 ksec, and 41 ksec, respectively.
The background spectra for these instruments were taken from Lockman-Hole observations
and subtracted before the spectral fitting below.
The EPIC data were corrected for telescope vignetting.
For spectral fitting, we utilize the SPEX (Kaastra et al. \cite{kaastra}) and XSPEC (Arnaud \cite{arnaud}) packages.
We model the plasma emission using the MEKAL emission code (Mewe et al. \cite{mewe}),
but with updated wavelengths according to Phillips et al. (\cite{phillips}).

\subsection{The EPIC results}
\begin{figure}
	\resizebox{\hsize}{!}{\includegraphics{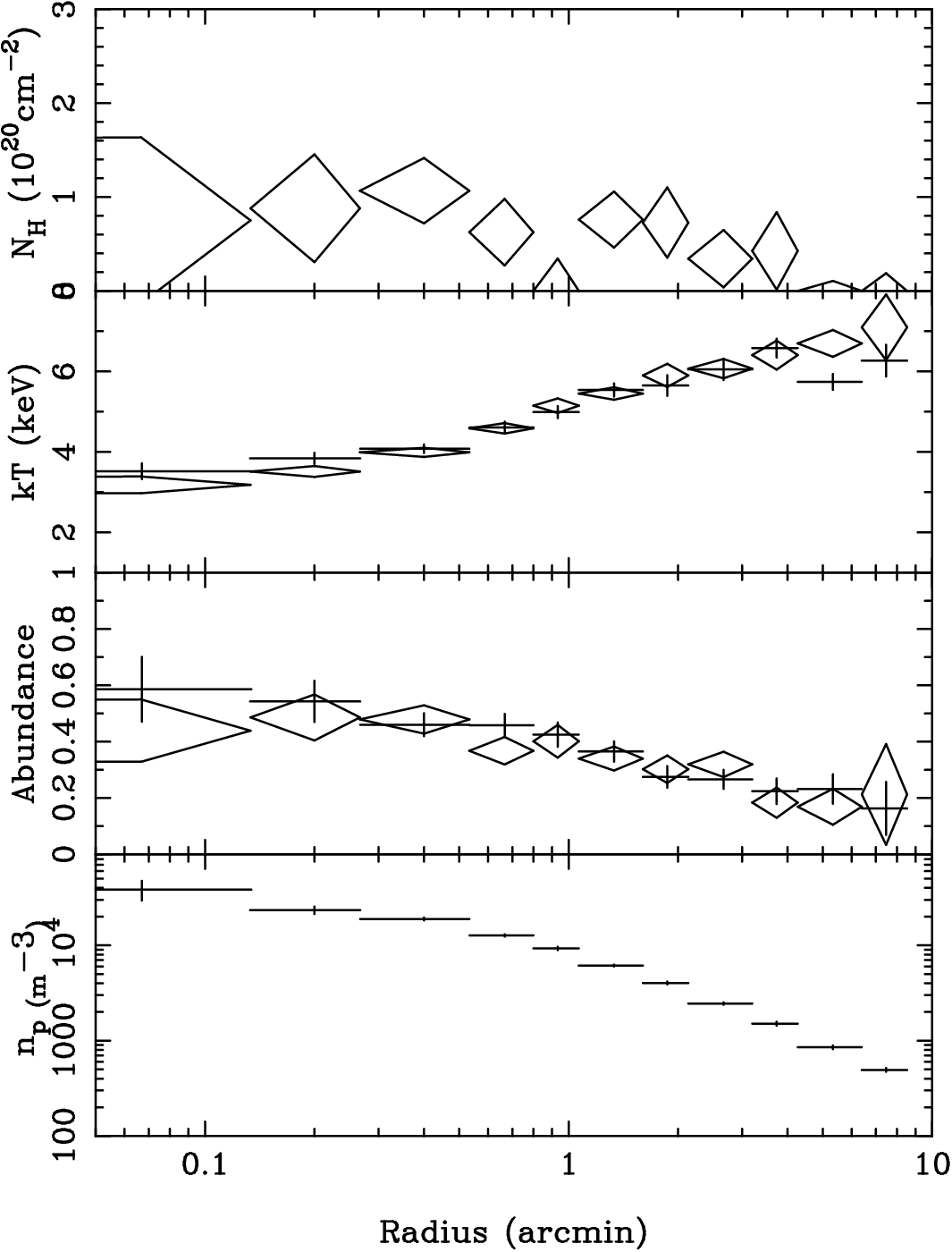}}
      \caption[]{Properties of the ICM as a function of projected radius
       derived from the PN and MOS spectra based on a single temperature model. 
       From top to bottom, absorbing column density, temperature, metal abundance, 
       and the deprojected hydrogen density were shown, respectively. 
       The PN and MOS results were shown by diamonds and crosses, respectively.}
         \label{fig:r-ta}
   \end{figure}

We extracted the spectra from the three EPIC detectors separately
in annuli around the emission center with outer radius ranging from $8''$ to $512''$.
The emission from the strongest point sources was subtracted from the spectra.
The MOS spectra were fitted with an isothermal model.
The column density $N_{\rm H}$ was assumed to be the Galactic value of $1.1\,10^{20}$~cm$^{-2}$.
We obtained statistically acceptable fits for all annuli.
The obtained temperature, metallicity, and hydrogen density
are shown in Fig.~\ref{fig:r-ta}.
The PN spectra were also fitted with the same model.
In this case, the $N_{\rm H}$ was left free, 
in order to constrain absorption due to possible cold gas in the cluster.
The fits are statistically acceptable for all annuli and the results are shown in Fig.~\ref{fig:r-ta}.

These results are consistent among the two instruments, except for temperatures measurements beyond 4\arcmin.
The difference in the temperature between PN and MOS may be caused by residual uncertainty
in the background subtraction.

The obtained column density from the PN indicates no excess absorption above the Galactic value.
Within $\sim$2\arcmin\ in radius from the center, 
the temperature decreases and the metallicity increases towards the center.
With the large effective area and moderate spatial resolution of the instrument,
we could resolve the ICM structure in sub-arcmin scale.

For further examination of the central temperature structure,
we have fit the EPIC spectra within a radius of 2\arcmin\ using two thermal components. 
The temperature of the hotter component was consistent with the value
obtained for the outer regions. 
For the cooler component we have obtained reasonable fits ($\chi^2/\nu$ of $380/282$ and $\sim 290/236$ for PN and MOS, respectively)
 using an isothermal component with a temperature of 2--4~keV. 
Alternatively, we attempted to model the cool component by an isobaric cooling flow model (Johnstone et al. \cite{johnstone}),
however in order to obtain a good fit to this model the temperature distribution had to be cut-off at $2.2-2.6$~keV. 
Essentially no cooler gas is seen directly in the EPIC data.
The mass deposition rate of the
cooling flow was poorly constrained (300--3000~\msun$/$year) due to a strong correlation
with the emission measure of the hot component. This is due to the relative high cut-off
temperature only a factor of 2--3 below the temperature of the hot component. 
Fits with a very small cut-off temperature did not produce acceptable fits with $\chi^2/\nu$ of $ \sim 425/284$ and $\sim 570/238$ for PN and MOS when $kT_{\rm min}=0.01$~keV.

\subsection{The RGS results}
The RGS spectra (RGS1 and RGS2) were extracted from the central $1'$ in full-width.
Because of the principle of the instrument, 
each spectrum contains emission from a range of different projected positions.
Based on the MOS image in the RGS energy band, 
we estimate that in the central RGS spectrum,
 more than about 50-60\% and 80\% 
of the flux comes from projected radii of $<30$\arcsec\ and $<60$\arcsec, respectively.
Based on these spectra, we examine the nature of the central region of the cluster.
We have detected the \ion{O}{viii} Lyman~$\alpha$ line clearly with an intensity of $1.2\,10^{51}$ photons/s.
The measured line wavelength is $20.184\pm0.03$~\AA\ and consistent with a redshift of 0.0631 within the statistical errors.
The wavelength uncertainty corresponds to a velocity shift of $\sim 500$ km~sec$^{-1}$.
The observed FWHM of the line is about 150 m\AA, which corresponds to a source extent of $\sim$1\arcmin\ in width.
In addition to the \ion{O}{viii} line, 
also some line blends in the wavelength range of 11--14~\AA\ are observed.
These are mainly due to \ion{Fe}{xxiv} (around 10.6~\AA\ and 11.2~\AA), \ion{Ne}{x} (12.1~\AA), and \ion{Fe}{xxiii} (12.2~\AA).
These line measurements should constrain the presence of material with a temperature of 0.5--3 keV.

\subsubsection{Isothermal model fitting}
The EPIC data showed that the cluster core region is close to being isothermal with a temperature of 3--4 keV.
For further examination of this isothermality, 
we fitted the central RGS spectra with an isothermal model.
The effects of a finite source extent were taken into account as follows.
We calculated the line spread function based upon the source surface brightness profile
and convolved this with the RGS response for a point source.
A $\beta$ model was assumed for the brightness profile.
The core radius was a free parameter in the spectral fitting procedure.
We limited the wavelength band to 10-23~\AA\ and 10-20~\AA\ for RGS1 and RGS2, respectively,
where most of the interesting emission lines are expected,
and the estimated background is less than 10\% of the source flux.

\begin{table}
\caption[]{The isothermal fits to the RGS spectra of the center of A1795. Numbers in parentheses are 90\% confidence uncertainties for one interesting parameter.}

\label{tbl:rgs-fits}
\begin{tabular}{lcc}
\hline 
Parameter	& RGS1	& RGS2 \\
\hline
$EM^a$		& 37	& 41 \\
$N_{\rm H}^b$	& 5.0 (3.5--6.4)	& 2.6 ($<4.6$) \\
$kT$ (keV)	& 3.8 (2.8--5.8)	& 4.0 (3.0--6.7) \\
O$^c$		& 0.30 (0.18--0.52)	& 0.30$^d$	\\
Ne$^c$		& 0.66 (0.31--1.4)	& 0.61 (0.37--1.4)\\
Fe$^c$		& 0.37 (0.21--0.85)	& 0.36 (0.21--1.1)\\
$\chi ^2/\nu$	& 356/310		& 222/247\\

\hline
\end{tabular}
\begin{description}
\item[$^a$] The volume emission measure in units of $10^{72}$m$^{-3} (10^{66}$cm$^{-3}$).
\item[$^b$] The column density in units of $10^{24}$m$^{-2} (10^{20}$~cm$^{-2}$).
\item[$^c$] Metal abundances relative to the solar values. The solar values are taken from Anders and Grevesse (\cite{anders}), with O/H$=8.51\,10^{-4}$, Ne/H$=1.23\,10^{-4}$, and Fe/H$=4.68\,10^{-4}$.
\item[$^d$] Due to the failure of one CCD on RGS2, the \ion{O}{viii} line was not detected with RGS2.
Therefore we fixed the O abundance for the RGS2 fit.
\end{description}
\end{table}

\begin{figure*}
	\resizebox{\hsize}{!}{\includegraphics[angle=-90]{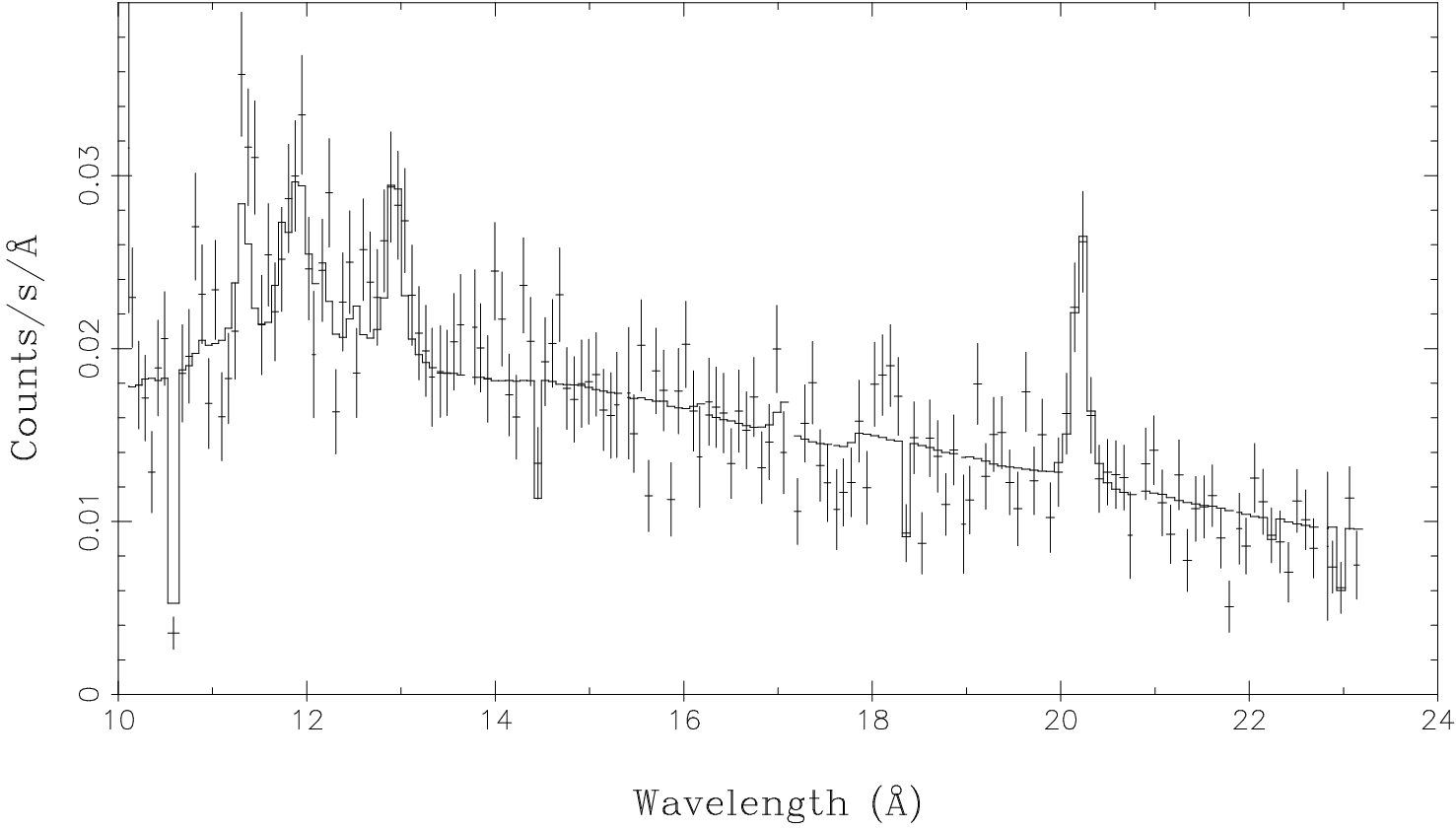}}
	\resizebox{\hsize}{!}{\includegraphics[angle=-90]{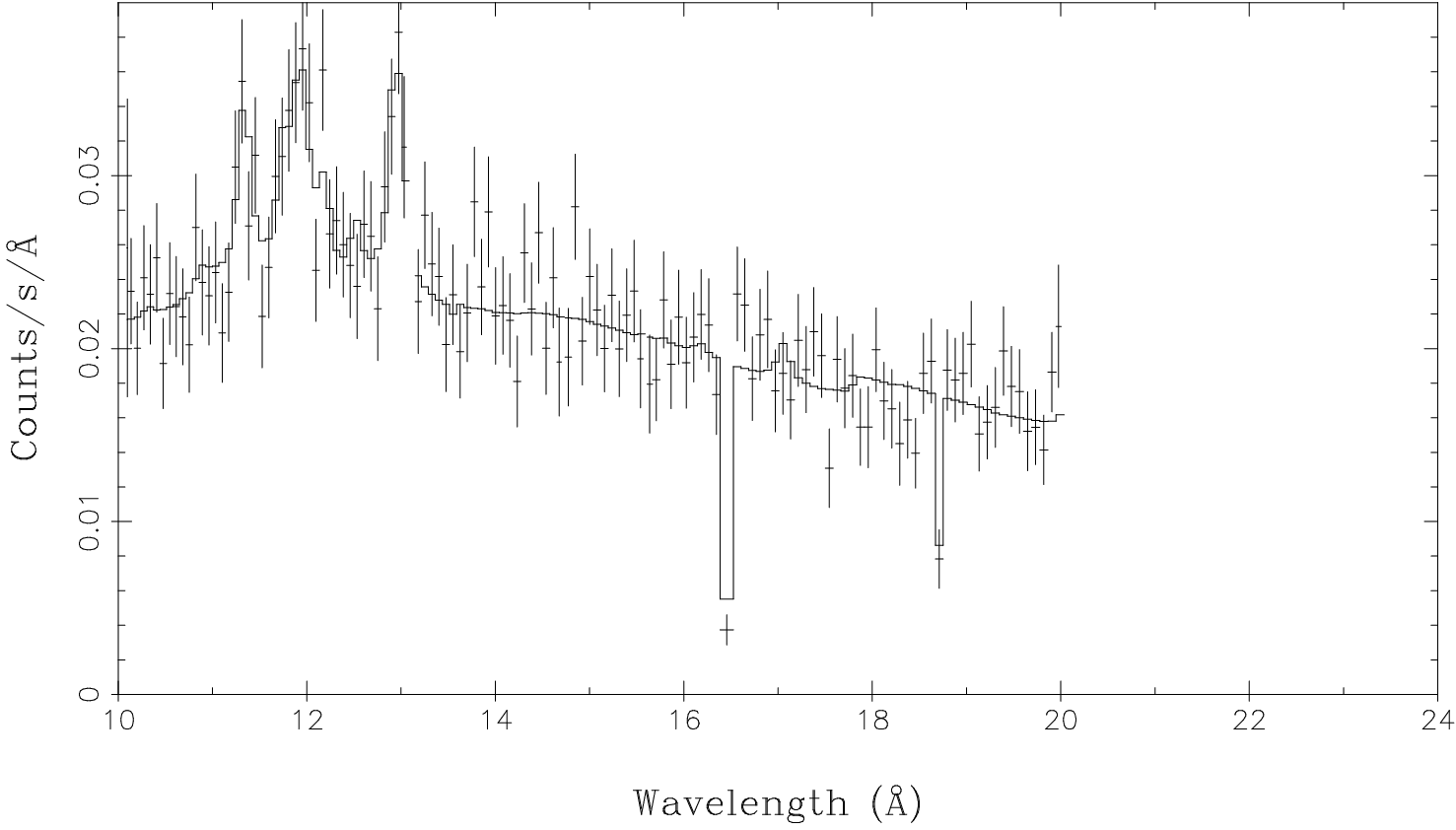}}
      \caption[]{The spectra extracted from the central region of A1795.
	The wavelength scales are not corrected for the redshift.
	Top and bottom panels show the RGS1 and RGS2 spectra, respectively.
	The data are shown with an isothermal model.
	Except for the normalization,	
	we plot the models with the average parameters from table.1.
	Due to CCD gaps and hot pixels, some wavelength bands were not covered by each RGS.
	These are indicated by gaps and drops in the model spectrum.
}
         \label{fig:rgs-spe}
   \end{figure*}

The fits to RGS1 and RGS2 are statistically acceptable and the best-fit parameters are shown in Table~\ref{tbl:rgs-fits}.
The best-fit parameters from the two detectors are consistent with each other.
The RGS spectra and the best-fit model are shown in Fig.~\ref{fig:rgs-spe}.

In the RGS1 spectrum,
there is line-like structure around 18~\AA\ in observed wavelength.
This could be due to line blends from \ion{Fe}{xvii}.
However, this structure is not seen in the RGS2 spectrum.
In addition, 
the maximum allowed emission measure from these blends, for a temperature of $\sim 0.5$~keV
 is on the order of 1\% of the total emission measure obtained above.
Therefore, we conclude this residual is not significant.
Non-statistical noise in background may be the origin of this structure.

The line profiles can be described using a core radius of 20\arcsec$-$30\arcsec\ for the assumed brightness profile.
This is roughly consistent with the observed X-ray brightness (Xu et al.  \cite{xu}; Allen \cite{allen}).

The $N_{\rm H}$ value obtained is above the Galactic value 
as well as the value obtained with the EPIC above.
However, the excess [$(2-3)\,10^{20}$~cm$^{-2}$] is within the systematic uncertainty due to the RGS response and the
estimated background.

\subsubsection{Limits on the cool component}
Thus, the RGS spectra do not require an additional cool component with a temperature lower than 3 keV 
in addition to the isothermal model.
Nevertheless, 
within the central region ($r<1'-2'$),
the cooling time inferred from X-ray observations
is much smaller than the possible cluster age (Edge et al. \cite{edge}).
This suggests the presence of cool material with a range of temperatures.
In fact, a recent Chandra observation revealed cool emission ($\sim 2$~keV) around the cD galaxy (Fabian et al. \cite{fabian-chandra}).

\begin{figure}
	\resizebox{\hsize}{!}{\includegraphics[angle=-90]{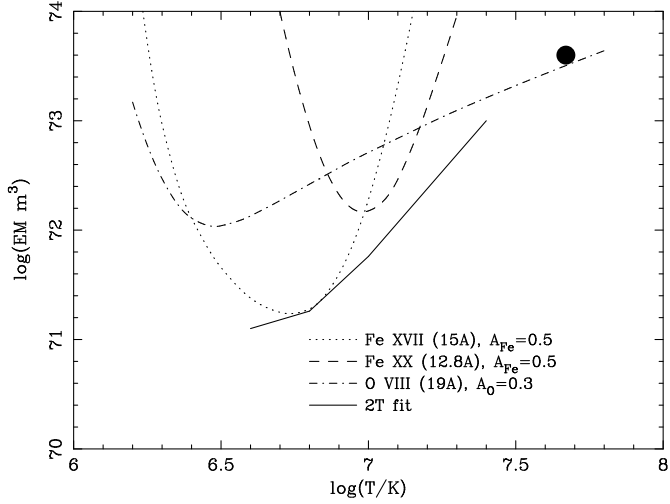}}
      \caption[]{The volume emission measure ($EM_{\rm cool}$) within a effective radius of 1\arcmin\ as a function of temperature $T_{\rm cool}$.
		Each line shows upper limits of $EM_{\rm cool}$ of the {\em isothermal} cool component
	 based on the upper limit of a line (and line blends) flux 
	and two temperature model fitting.
	The filled circle indicates the emission measure and temperature of
	 the hot component derived from the isothermal model fit to the RGS spectra. }
        \label{fig:ei-t}
   \end{figure}

In order to constrain the possible cool component,
we fitted the data by adding a cooler plasma component in addition to the isothermal model.
Firstly, we considered an additional isothermal component.
Since the RGS is not so sensitive to hot  plasma ($>3$ keV)
we fixed the hot component temperature $T_{\rm hot}$ to 6.4 keV, the ICM value obtained with the EPIC.
The abundances of both components were fixed to 0.4 solar, except for the Ne abundance.
The Ne abundance was fixed to 0.64 times solar based on the isothermal fits.
The $N_{\rm H}$ value was also fixed to $3\,10^{20}$~cm$^{-2}$, which is the best-fit value for the isothermal model.
We measured the upper limits to the volume emission measure of the cool component, $EM_{\rm cool}$,
for a given temperature of the cool component, $T_{\rm cool}$.
The emission measure of the hot component, $EM_{\rm hot}$, was left free.
The results are shown in Fig.~\ref{fig:ei-t}.
When $T_{\rm cool}$ is below 1.5~keV, 
$EM_{\rm cool}$ is less than 10\% of $EM_{\rm hot}$.

   \begin{figure}
	\resizebox{\hsize}{!}{\includegraphics[angle=-90]{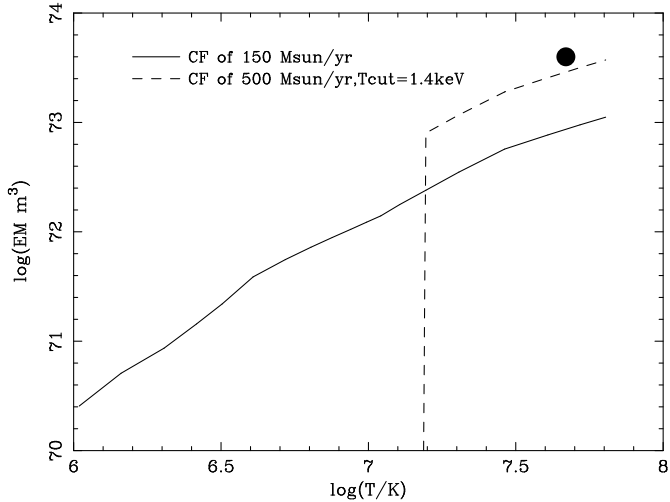}}
      \caption[]{The cumulative volume emission measure [$EM_{\rm cool}(<T)$] as a function of temperature. Two lines are based on cooling flow models which reproduce the RGS spectra.
			}
         \label{fig:ei-cf}
   \end{figure}
Secondly, 
we considered a continuous distribution in temperature
based on an isobaric cooling flow model (CF; Johnstone et al. \cite{johnstone}).
In this model, 
$EM_{\rm cool}$ was assumed to be
\[ \int dEM(T) = \frac{5}{2}\frac{\dot{M}}{\mu m_{\rm p}} \int^{T_{\rm max}}_{T_{\rm min}} \frac{dT}{\Lambda(T)} ,\]
where $\dot{M}, \mu,$ and $m_{\rm p}$ 
are the mass deposition rate, mean molecular weight, and the proton mass, respectively.
$\Lambda(T)$ is the cooling function.
The temperature of the hot component, $T_{\rm hot}$,  
the abundances of both components, and $N_{\rm H}$ were fixed to the same values as above.
The maximum temperature of the CF, $T_{\rm max}$, was fixed to the same value as $T_{\rm hot}$.
This CF component was added to the isothermal component.
We found a 90\% upper limit to $\dot{M}$ of 150 \msun$/$yr.
This is significantly smaller than the result based on the deprojection analysis (Edge et al. \cite{edge}) where a cooling flow age of a Hubble time was assumed.
When we assume a cooling flow with $\dot{M}=500$ \msun$/$yr based on Edge et al. (\cite{edge}), 
the lower temperature should be larger than $\sim$  1.4 keV in order to produce
the observed RGS spectra.
The cumulative $EM_{\rm cool}$  of the CF model with this upper limit $\dot{M}$ is shown in Fig.~\ref{fig:ei-cf}.
Similar to the two temperature model fitting,
the emission measure of the cool component is very small compared to that of the hot component.

These tight upper limits were obtained through the lack of emission lines 
such as  \ion{Fe}{xx} (12.8~\AA\ in the rest frame) and \ion{Fe}{xvii} (around 15~\AA\ and 17~\AA) in the observed spectra.
In order to constrain the limit more directly and confidently,
we focused on particular ionization stages of ions and their line emissions.

The \ion{Fe}{xvii} line at a rest wavelength of 15.0~\AA\ is a sensitive indicator of plasma with a temperature of 0.2--0.8 keV.
As shown in Fig.~\ref{fig:rgs-spe}, 
the RGS spectrum indicates no significant residual at this wavelength.
By adding this line to the isothermal model
we derived an upper limit to the line intensity of $2\,10^{50}$ photons$/$s.
A further constraint can be derived from the \ion{Fe}{xx} blends at a rest frame wavelength of 12.8~\AA.
This is sensitive to a temperature of 0.5--2.0 keV.
The upper limit to the \ion{Fe}{xx} blends is $5\,10^{50}$ photons$/$s.

These line measurements place limits on the presence of cool component(s)
without the assumptions used above on the temperature structure and the detailed emission model.
We calculate $EM_{\rm cool}$ based on the line intensities measured for a given $T_{\rm cool}$ and metallicity, 
following Canizares et al. (\cite{canizares}).
The result is shown in Fig.~\ref{fig:ei-t}.
The predicted emission measure based on the line intensity of the \ion{O}{viii} line
is also shown, for reference.
Here we assumed an O and Fe abundance of 0.3 and 0.5 solar, respectively, based on our measurement.
The emission measure of the possible cool component is less than 10\% of that of the hot component
within the central volume
when $T_{\rm cool}$ is below 1 keV.

When we assumed the isobaric cooling flow model for $EM_{\rm cool}$,
the limits for \ion{Fe}{xvii} correspond to $\dot{M}$ = 120~\msun$/$yr.
This is similar to the value obtained above based on the spectral fitting, 
indicating that $\dot{M}$ is strongly constrained by a lack of the \ion{Fe}{xvii} lines in the spectra.

\subsubsection{Oxygen abundance}
The abundance of O and its ratio to other elements in the ICM strongly depends on its origin.
Nevertheless, previous measurements of the O abundance had a large uncertainty (e.g. Canizares et al.  \cite{canizares}), 
except for a few cases (e.g., Matsumoto et al.  \cite{matsumoto}).
This is not only because of the limited sensitivity to O lines
but also because of uncertainty in the temperature structure of the ICM.
We detected the \ion{O}{viii} Lyman $\alpha$ line clearly for the first time in this cluster
and we constrained the temperature structure tightly. 
This provides a good opportunity to measure the O abundance.

As shown in Table~\ref{tbl:rgs-fits}, the isothermal fits to the RGS data yielded an O abundance of 0.2--0.5 with respect to solar.
When we used the central ICM temperature of 4.5 keV determined by EPIC,
the O abundance is 0.27--0.47.

\section{Summary and Discussion}
Based on the XMM observations of Abell 1795, 
we have obtained the following results.

The temperature and metallicity of the ICM
are radially uniform at radii from 2\arcmin--8\arcmin\ ($\sim 200-800$ kpc),
with an observed temperature of 6--7 keV and metallicity of 0.2--0.3 solar.
This confirms the previous idea that the cluster is close to being dynamically relaxed (e.g., Briel and Henry \cite{briel}).

Below a radius of $\sim 2'$ the effective temperature of the ICM starts to decrease towards the center
from 6--7 keV to 3--4 keV.
Except inside the central 1\arcmin\, the observed temperature profile agrees well 
with that obtained from a deprojection analysis based on the cooling flow model (Edge et al. \cite{edge}).

On the contrary, 
we found lack of emission from cool ($<1-2$~keV) gas in the cluster center, 
compared to the prediction from the strong cooling flow ($\dot{M} \sim 500$~\msun$/$yr).
The EPIC and RGS spectra extracted from the cluster center ($r<30$\arcsec$-60$\arcsec)
can be described by an isothermal model with a temperature of $\sim 4$ keV.
Furthermore, through the absence of clear \ion{Fe}{xvii} and \ion{Fe}{xx} line emission in the RGS spectra, 
the emission measure of any cool component with a temperature in the range $0.4-1.3$~keV 
is found to be at least a factor of 3 lower than that would be expected from the strong cooling flow.
The absence of the predicted amount of cool emission is also supported by the cut-off temperature of $\sim$2.4~keV found by
 fitting cooling flow models to the EPIC data.
The absence of these emission lines is difficult to model with a uniformly covering absorber,
both covering the hot and cool components, 
and we derive a covering fraction of $<20$\% 
for gas of sufficient column density ($10^{21}$~cm$^{-2}$)
necessary to ``hide'' the iron lines.
We derived an upper-limit to $\dot{M}$, of $150$ \msun$/$yr,
which is smaller than found by Allen et al. (\cite{allenetal}; $250\pm 70$ ~\msun$/$yr) based on the ASCA spectral analysis.

The present observation indicates that at temperatures greater than 3 keV the ICM is cooling as predicted by the cooling flow picture. 
Nevertheless, we have no direct spectral evidence that the gas continues cooling below 2 keV.

This result for A~1795 is similar to what has been found in A~1835 (Peterson
et al. \cite{peterson}) and S\'ersic 159-03 (Kaastra et al. \cite{sersic}).
A full discussion of possible explanations for the lack of observed cool gas is
given by Peterson et al. (\cite{peterson}) in their analysis of the A~1835 cluster.

We detected the \ion{O}{viii} Ly~$\alpha$ line from the cluster center with the RGS.
The O abundance was found to be 0.2--0.5 solar value.
This is similar to the value obtained in the center of the Virgo cluster (Matsumoto et al. \cite{matsumoto}) based on ASCA spectra, 
and in the centers of A~1835 (Peterson et al. \cite{peterson}) and S\'ersic 159-03 (Kaastra et al. \cite{sersic}) based on XMM/RGS spectra.
The O to Fe ratio at the cluster center is $0.5-1.5$ times the solar ratio.
Since oxygen and iron are mainly produced in type II and Type Ia supernovae, respectively,
the observed ratio suggests significant contributions from both type Ia and type II  to the metal enrichment at the cluster center.
In the core the Fe abundance is a factor of 2 larger than in the outer parts (see Fig.~\ref{fig:r-ta}).
This is consistent with the general idea that the iron abundance is higher around the
cD galaxy (Fukazawa \cite{fukazawa-phd}; Fukazawa et al. \cite{fukazawa}).

\begin{acknowledgements}
This work is based on observations obtained with XMM-Newton, an ESA science 
mission with instruments and contributions directly funded by 
ESA Member States and the USA (NASA).
We thank Dr.E.Churazov for valuable comments.
The Laboratory for Space Research Utrecht is supported
financially by NWO, the Netherlands Organization for Scientific
Research. 
\end{acknowledgements}

\end{document}